\documentclass[a4paper, accepted=2020-12-31]{quantumarticle}
\pdfoutput=1
\usepackage[numbers,sort&compress]{natbib}
\usepackage{hyperref}
\usepackage[utf8]{inputenc}
\usepackage[english]{babel}
\usepackage[T1]{fontenc}
\usepackage{amsmath}
\usepackage{amsfonts}
\usepackage{amssymb}
\usepackage{graphicx}
\usepackage{xcolor-solarized}
\usepackage{amsthm}
\usepackage{color}

\usepackage{textcomp}

\newcommand{\real}{\mathbb{R}}

\newcommand{\aref}[1]{\hyperref[#1]{Appendix~\ref{#1}}}

\graphicspath{{figures/}}

\begin{document}
	
\title{Neural Network Decoders for Large-Distance 2D Toric Codes}

\author{Xiaotong Ni}
\affiliation{QuTech, Delft University of Technology, P.O.Box 5046, 2600 GA Delft, The Netherlands.}
\email{xiaotong.ni@gmail.com}

\date{2020 April 8}

\maketitle

We still do not have perfect decoders for topological codes that can satisfy all needs of different experimental setups.
Recently, a few neural network based decoders have been studied, with the motivation that they can adapt to a wide range of noise models, and can easily run on dedicated chips without a full-fledged computer.
The later feature might lead to fast speed and the ability to operate at low temperatures.
However, a question which has not been addressed in previous works is whether neural network decoders can handle 2D topological codes with large distances.
In this work, we provide a positive answer for the toric code~\cite{Kitaev2003Faulttolerantanyon}.
The structure of our neural network decoder is inspired by the renormalization group decoder~\cite{duclos2010fast,duclos2013fault}.
With a fairly strict policy on training time, when the bit-flip error rate is lower than $9\%$ and syndrome extraction is perfect, the neural network decoder performs better when code distance increases.
With a less strict policy, we find it is not hard for the neural decoder to achieve a performance close to the minimum-weight perfect matching algorithm.
The numerical simulation is done up to code distance $d=64$.
Last but not least, we describe and analyze a few failed approaches.
They guide us to the final design of our neural decoder, but also serve as a caution when we gauge the versatility of stock deep neural networks.
The source code of our neural decoder can be found at~\cite{sourcecodegithub}.

\section{Introduction}
\label{sec:Intro}
Before we can make the components of quantum computers as reliable as those of classical computers, we will need quantum error correction so that we can scale up the computation.
The surface code and other topological codes are popular choices for several qubit architectures because of their high thresholds and low requirement on connectivity between qubits.
However, several good performing decoders have trouble to do real-time decoding for qubits with fast error-correction cycles, such as superconducting qubits.
Moreover, as we are getting closer to the point where small size surface code can be implemented in the lab, it is desirable that the decoders can adapt to the noise models from the experimental setups.
These considerations motivate the study of decoders based on neural networks, which we will refer to as neural decoders, for surface code and other topological codes~\cite{Baireuther2018Machinelearningassisted,Varsamopoulos2017Decodingsmallsurface,torlai2016neural,Breuckmann2018Scalable,Baireuther2018Neuralcolorcode,Krastanov2017neural,Maskara2019advantages,Chamberland2018neural}.
One question that has not been addressed so far is whether neural networks can also be used for decoding 2D topological codes on a large lattice with good performance.
In this work, we will focus on answering this question for the toric code.
While it is the simplest topological code, it shares many common features with others, which makes it a good test platform.

To design a neural decoder for large toric codes, a natural first step is to use convolutional neural networks (CNNs)~\cite{lecun1989generalization,Krizhevsky2012imagenet}, as the toric code and CNNs are both translational-invariant on a 2D-lattice.
Compared to normal neural networks, the number of parameters in CNNs only scale with the depth of networks.
This gives an intuition that the training of the CNNs should remain feasible for the lattice size of concern in the near future.
We want the decoder to be able to adapt to experimental noise, which we should assume to be constantly changing, and thus the data for calibration is limited.
The structure of CNNs allows us to have a great control of how many parameters to be re-trained during calibration so that we can avoid over-fitting (see \aref{appendix:varying_perror} for an example).

Interestingly, the renormalization group (RG) decoder~\cite{duclos2010fast,duclos2013fault} for toric code already has a structure very similar to the CNNs used in image classification.
Both of them try to keep the information needed for the output intact while reducing the size of the lattice, by alternating between local computation and coarse-grain steps.
This similarity means that we should aim to achieve better or similar performance with the neural decoder compared to the RG one.
And in case of bad performance, we can ``teach'' the neural decoder to use a similar strategy as the RG decoder.
This is indeed how we get good performance in the end.
Conceptually, this is similar to imitation learning (see~\cite{Attia2018Globaloverviewof} for an overview).
Even though we initialize the neural decoder by mimicking the RG one, it can have the following advantages:
\begin{itemize}
	\item It can achieve a better performance than the RG decoder, as the latter one contains some heuristic steps. On the other hand, the neural decoder can be optimized to be a local minimum with respect to the parameters of the neural network (strictly speaking, at least the gradient is very small).
	The idea of improving belief propagation with neural networks is also used for decoding classical linear codes~\cite{Nachmani2016Learningtodecode}.
	\item It offers an additional way to adapt to experimental noise models, which is simply training on experimental data.
\end{itemize}

It is tricky to evaluate the performance of neural decoders.
As it stands, we need to train the neural nets for different lattice sizes separately, and the training process is not deterministic.
Thus, we cannot define a threshold for the decoder.
This is fine if our main goal is to have an adaptable decoder for near-future quantum devices.
However, in order to know how optimal the neural decoders are, we still make a ``threshold plot'' under a well-studied noise model in \autoref{fig:logical_vs_physical}.
Roughly speaking, the threshold benchmark is a good indicator of how good the decoder can process syndrome information.
At the same time, we compare our neural decoder to the minimum-weight perfect matching algorithm in \autoref{fig:comparison_to_MWPM}, and show in \aref{appendix:varying_perror} that our neural decoder can improve itself when trained on different error model.
We hope these pieces of information together can give us a first impression of neural decoders on toric code.

The focus of this paper is not on how to obtain an optimal neural decoder.
Indeed, a lot of hyper-parameter optimizations can be done to further improve the performance or reduce the amount of data needed for training.
Instead, we describe the key ideas that allow us to reliably obtain decent neural decoders for the toric code.
The knowledge we gained can help us design neural decoders for other large codes.

\section{Introduction to Toric code and the Renormalization Group Decoder}
\label{sec:toric_code_and_rn_decoder}
\subsection{Toric Code}

\begin{figure}
	\centering
	\includegraphics[width=0.5\linewidth]{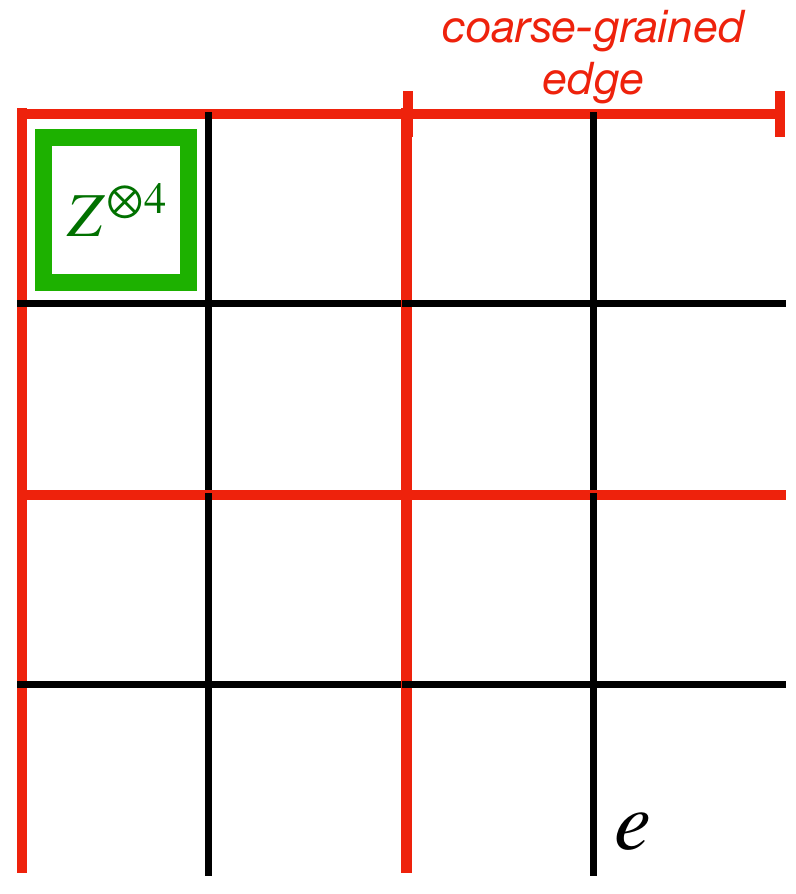}
	\caption{An illustration of the lattice and a $B_p$ stabilizer check. The $2\times 2$ unit cells are marked by the red color. We also give an example of a coarse-grained edge, which locates at the top-right and contains two edges of the original lattice.}
	\label{fig:lattice}
\end{figure}

First, we give a brief introduction of toric code.
Consider a $L\times L$ square lattice with periodic boundary condition, where a qubit lives on each edge.
The stabilizer group of toric code is generated by two types of operator $A_s$ and $B_p$
\begin{equation}
A_s = \bigotimes_{q\in n(s)} X_q, \quad B_p = \bigotimes_{q\in n(p)} Z_q,
\end{equation}
where $s$ and $p$ is any site and plaquette respectively, and $n(\cdot)$ consists of the 4 qubits neighboring $s$ or $p$.
The logical-$Z$ operators have the form
\begin{equation}
\bar{Z}_i = \bigotimes_{q\in l_i} Z_q ,
\end{equation}
where $l_{1,2}$ are two shortest inequivalent non-contractible loops.
The toric code has a distance $d=L$.

In this paper, we will focus on the bit-flip noise model, i.e. only $X$ errors can happen.
We will also assume perfect measurements.
Under this restriction, the quantum states will stay in the $+1$ eigenspace of $A_s$.
Therefore, we only need to consider the expectation values of $B_p$ and $\bar{Z}_i$.
For simplicity, let us suppose in the beginning $\langle \bar{Z}_i \rangle = +1$.
And then a set of $X$ errors happened, which leads to the syndrome $s=\{\langle B_p \rangle \}$.
The goal of a decoder is to apply $X$ to the qubits, such that $\langle B_p \rangle$ and $\langle \bar{Z}_i \rangle$ return to $+1$.
Without going to detail, we claim it is enough to know the parity of the number of $X$ errors that happened on the loops $l_{1,2}$.
These two parities will be the final training target for our neural decoder.
We will refer to the two parities as logical corrections.

\subsection{Renormalization Group Decoder}

Let us first set up some notation.
We will use $e$ to denote an edge of the original lattice or a coarse-grained edge.
When we say $e$ is a coarse-grained edge, we mean $e$ is an edge of a unit cell which consists of two (or more) edges of the original lattice.
We use $x(e) = 1$ to denote an $X$ error happened on edge $e$, and otherwise $x(e)=0$.
When $e$ is a coarse-grained edge consists of edges $\{e_i\}$, we set $x(e)$ to be 
\begin{equation}
x(e)= \sum_i x(e_i) \mod 2.
\end{equation}
Lastly, the conditional marginal probability distribution $p_e\left(x\left(e\right)| \text{syndrome}\right)$ of error on a coarse-grained edge $e$ is denoted by $p_e$.
Theoretically, $p_e$ can be computed by enumerating all error configurations that have the given syndrome.
However, with renormalization group decoders and our neural decoders, we will only be able to compute approximate distributions $p'_e \approx p_e$.
Therefore, with a slight abuse of notation, we will use $p'_e$ and $p_e$ interchangeably.
We will also use $p_e$ to denote the physical error rate of an original edge $e$.

\begin{figure}
	\centering
	\includegraphics[width=0.9\linewidth]{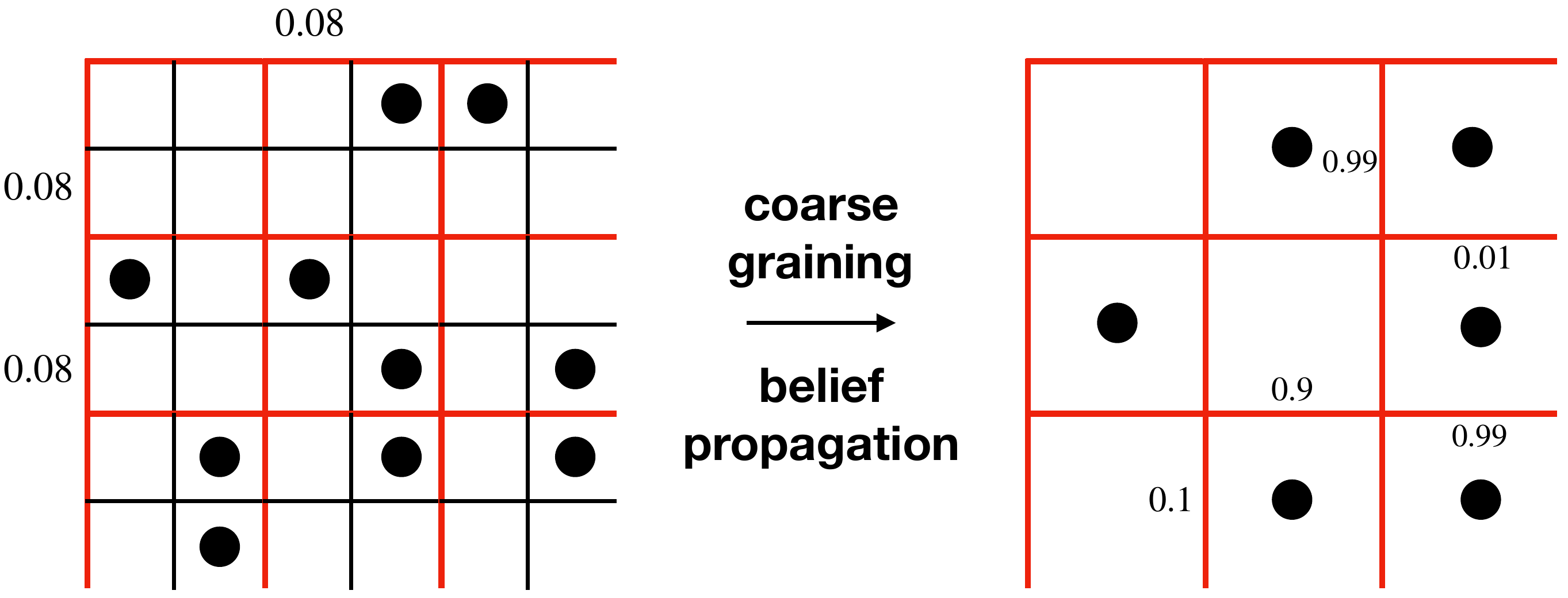}
	\caption{An illustration of the first renormalization stage. The numbers in the figure are made up by the author and are likely not accurate. The left side is the physical lattice, where the qubits have the same error rate. The dots represent the locations of violated parity checks. The right side is the coarse-grained lattice. The dots indicate that the corresponding coarse-grained plaquettes contain an odd number of violated parity checks. The numbers on coarse-grained edges are $\{p'_e(1)\}$, which are computed by belief propagation. With the help of $\{p'_e(1)\}$, we can more reliably figure out how the coarse-grained syndrome should be matched.}
	\label{fig:bp_example}
\end{figure}

One renormalization stage consists of the following (see \autoref{fig:bp_example} for an example):
\begin{enumerate}
	\item Divide the lattice into $m\times m'$ unit cells, where in this work $m=m'=2$.
	\item The ideal outputs of the renormalization step are $\{p_e\}$ for each coarse-grained edge $e$ that is a border of a unit cell.
	However, we can only compute $\{p_e\}$ approximately by using belief propagation, which is a heuristic procedure for computing marginal probabilities (see \aref{appendix:bp}).
	These approximate $\{p'_e\}$ are treated as the error rate of the coarse-grained edge $e$ for the next renormalization stage.
\end{enumerate}

At the end of the renormalization process, we obtain $p_e$ for $e$ being either of the two non-contractible loops.
For simplicity, we assume the two non-contractible loops are $l_{1,2}$.
Thus, we get an approximation of the marginal probability distribution for logical correction.

\section{Design and Training of the Neural nets}
\label{sec:design}
At a first glance, to build a neural decoder, we can simply train a convolutional neural net with input-output pairs (syndrome, logical correction).
However, in practice, this does not allow us to get a good enough performance.
A detailed description of some simpler approaches and discussion will be presented in \aref{appendix:simpler_approach}.
Those failures eventually motivate us to design and train the neural decoder in the following way.

\subsection{Design of the network}
\label{subsec:design_network}
\begin{figure}
	\centering
	\includegraphics[width=.8\linewidth]{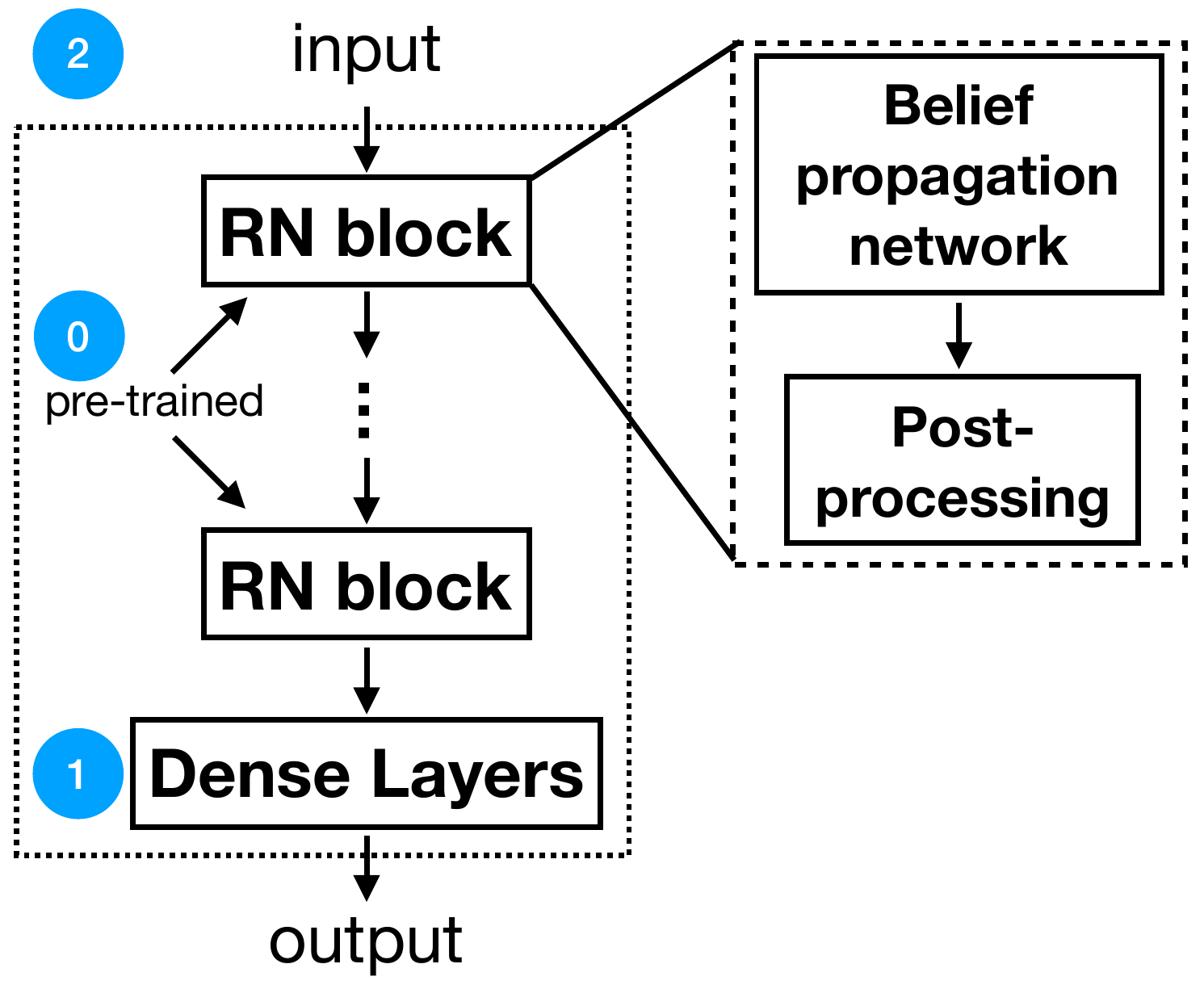}
	\caption{Structure of the entire network, and the training order labeled by the blue circles. After loading the pre-trained belief propagation network into RN block, the first step is to train the dense layers, and the second step is to train all the layers together. We will call the second step global training.}
	\label{subfig:complete_network}

\end{figure}

The network follows the same structure as the renormalization decoder.
Most of the network is repetitively applying the renormalization block, which is depicted in \autoref{subfig:complete_network}.
The belief propagation (BP) network, as its name suggested, is intended to approximate the BP process (see \aref{appendix:bp} for an introduction).
More concretely, the first step of the training process is to train the BP network with the data generated by a handcrafted BP algorithm.
This means initially the inputs to the BP network are syndromes and error rates on each edge, and the outputs are supposed to approximate the error rates on coarse-grained edges.
However, later in the training process (i.e. global training mentioned in \autoref{subsec:training}), the BP network can deviate from this initial behavior.
The post-processing has two steps.
The first step is to remove the superficial complexity from the coarse-grained lattice.
Whenever for a coarse-grained edge $e$ has $p_e(1)>0.5$, we apply an $X$ on $e$ and switch $p_e(1) \leftrightarrow p_e(0)$.
If $e$ is on either of the two non-contractible loops $l_{1,2}$, then the desired logical correction will be updated accordingly.
Although this step only changes the representation of the data, and in principle, neural nets can learn to do the same thing, it is a quite costly step for neural nets as it can call the parity function multiple times.
The second step is coarse-graining.
We need to reduce the lattice size by half, and for convenience, this is done by the first layer of every belief propagation network.
We also compute the parity of four $\langle B_p \rangle$ in each unit cell and feed these parities to the next BP network as the effective syndrome of the coarse-grained lattice.

In more detail, the input to the BP network can be packed in a tensor $I$ with shape $(l,l,3)$, where $l$ is the initial lattice size or the output size of the precedent BP network.
For example, we can set $I(i,j,0)$ to be $\langle B_p \rangle$ on plaquette $(i,j)$, and $I(i,j,1), I(i,j,2)$ to be the error rates corresponding to the top and left qubits of the plaquette.
Each BP network consists of 13 convolution and 3 batch normalization layers.
The definition of convolution layers can be found in \aref{appendix:intro_nn}, and batch normalization is introduced in~\cite{Ioffe2015BatchNormalization}.
The first layer reduces the lattice size $L$ by half.
The reasoning is that the belief propagation is done based on $2\times 2$ unit cells.
The remaining layers keep the lattice size unchanged.
Among them, only four involve communication between unit cells, i.e. the kernels of these four convolution layers have size $3\times 3$.
They spread evenly in the 13-layer network.
Other layers only have kernels of size $1\times 1$, which can then be viewed as computation inside unit cells.
The rationale behind this is that the messages likely need to be processed before the next round of communication.
The batch normalization layers also spread evenly, with the hope that they can make the training more stable.

After the renormalization process reduces the lattice to a size of $2\times 2$, we apply 4 dense layers (a.k.a fully-connected layers).
Note that the dense layers conveniently break the translational symmetry imposed by the convolution layers.
In the end, we have a neural network with input shape $(L,L,3)$ and output shape $(2)$\footnote{For efficient training, an additional dimension called batch size will be added.}.
The input shape is $(L,L,3)$ because this is the input shape of BP networks.
For $L=64$, the total number of trainable layers in the network is around 60, which is very large compared to early deep neural networks~\cite{Krizhevsky2012imagenet}.
However, most of the computation cost and the trainable parameters are concentrated in the 16 convolution layers with kernel size $3\times 3$.
Combining this and the careful training strategy we describe below, we find that the training can be done very efficiently.

\subsection{Training}
\label{subsec:training}
In general, training neural networks becomes harder when the number of layers increases.
This is often attributed to the instability of gradient backpropagation.
Considering we have a very deep neural network, we should find a way to train parts of the network first.
The training is divided into two stages.
First, we train the belief propagation network to indeed do belief propagation (BP).
This corresponding to the blue circle with 0 in \autoref{subfig:complete_network}.
To do this, we implement a BP algorithm and use it to generate training data for the network.
More concretely, we first assign a random error rate $e^{-k}$ to each edge, where $k\in [0.7,7]$ from a uniform distribution.
The choice of the distribution is quite arbitrary.
Then we sample error on each edge according to its error rate and compute the syndrome.
After that, we feed both the error rates and syndrome into our handcrafted BP algorithm, which will output an estimation of the error rates $p_e$ corresponding to the coarse-grained edges.
We can subsequently train the BP network with the same input-output relation.
An important detail is that we transform the error rates $p_e(1)$ in both input and output to $r_e=\log \left(p_e(1)/p_e(0)\right)$.
The reason behind this is described in \aref{appendix:simpler_approach}. 

Next, we load the pre-trained belief propagation network into the decoder network described in the previous subsection.
To ensure $r_e$ stay bounded, we perform a rescale $r_e \rightarrow 7r_e/\max_e |r_e|$ before feed it into next RN block (the choice of 7 here is arbitrary).
We can then train the dense layers and afterward the whole network with input-output pairs (syndrome, logical correction).
These two trainings correspond to the blue circle 1 and 2 in \autoref{subfig:complete_network}, respectively.
The training data is measurable in experiments in these two training.

We train the decoders for different lattice sizes $L$ separately.
Although this makes the concept of threshold pointless, it is still useful to estimate the ``threshold'' so that we can have a rough comparison of the neural decoder with the existing ones.
For this, we train the decoder for different $L$ with the same amount of stochastic gradient steps, which also implies the optimizer sees the same amount of training data for each $L$.
In addition, the training for each $L$ is done under 1 hour (on the year 2016 personal computer with 1 GPU).
We consider this to be a fairly strict policy.
The result is plotted in \autoref{fig:logical_vs_physical}.
We can also forgo this strict policy and spend more time in training the neural decoder for $d=64$ toric code, which gives rise to \autoref{fig:comparison_to_MWPM}.
The training time is still under 2 hours.
More details about the design and training can be found in \aref{appendix:details} and the source code~\cite{sourcecodegithub}, and more discussion about the numerical results can be found in the following section.

\section{Numerical results}
\label{sec:numerics}
\begin{figure}
	\centering
	\includegraphics[width=\linewidth]{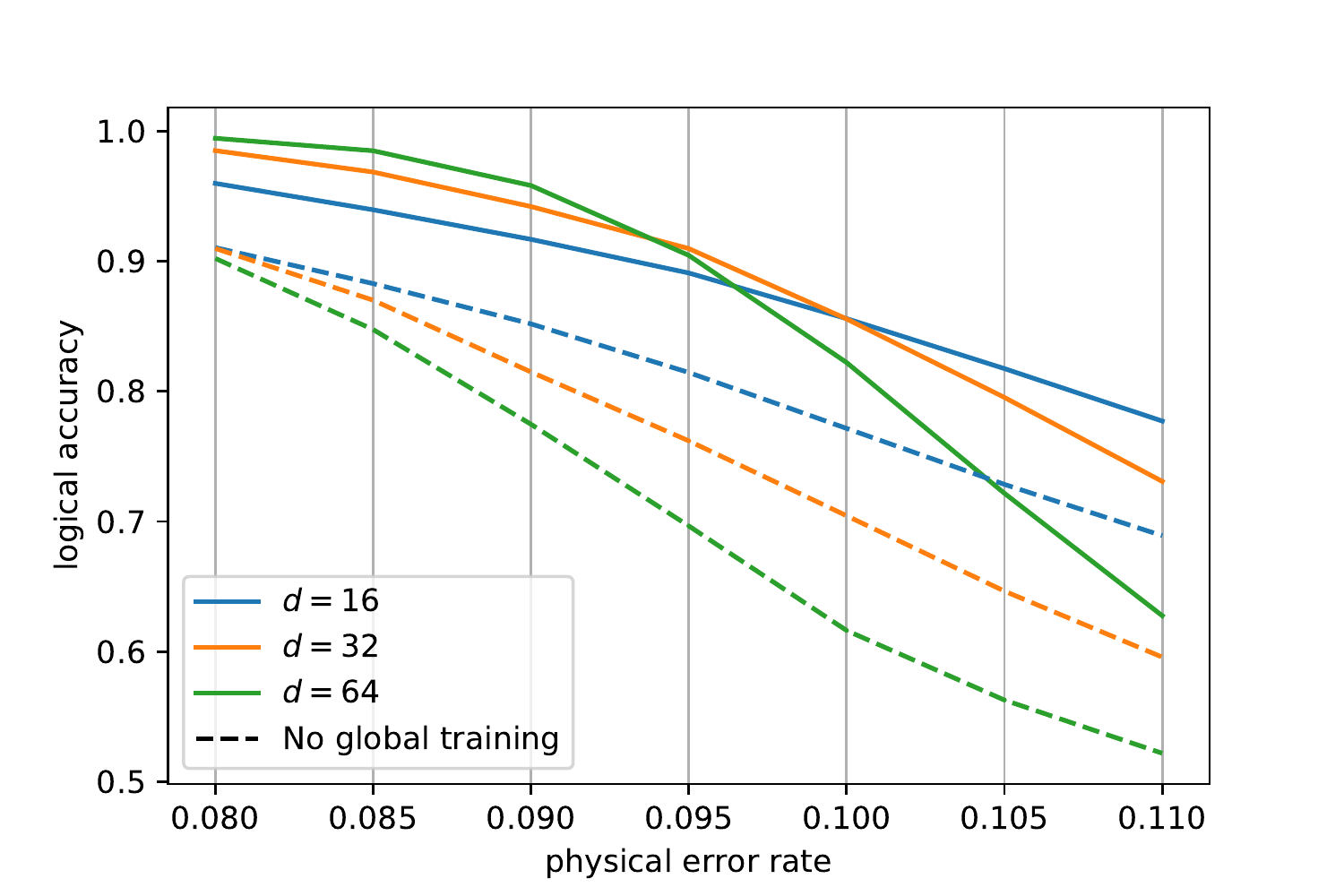}
	\caption{Logical accuracy versus physical error rate.
		The neural decoders are trained at physical error rate $9\%$.
		For the three solid lines, the decoder has been trained globally, while the dashed lines it has not.
		The colors of the dashed lines indicate the code distance they are evaluated on.
		The vertical grid indicates the physical error rates for which we evaluate the logical accuracy.
	}
\label{fig:logical_vs_physical}
\end{figure}

\begin{figure}
	\centering
	\includegraphics[width=\linewidth]{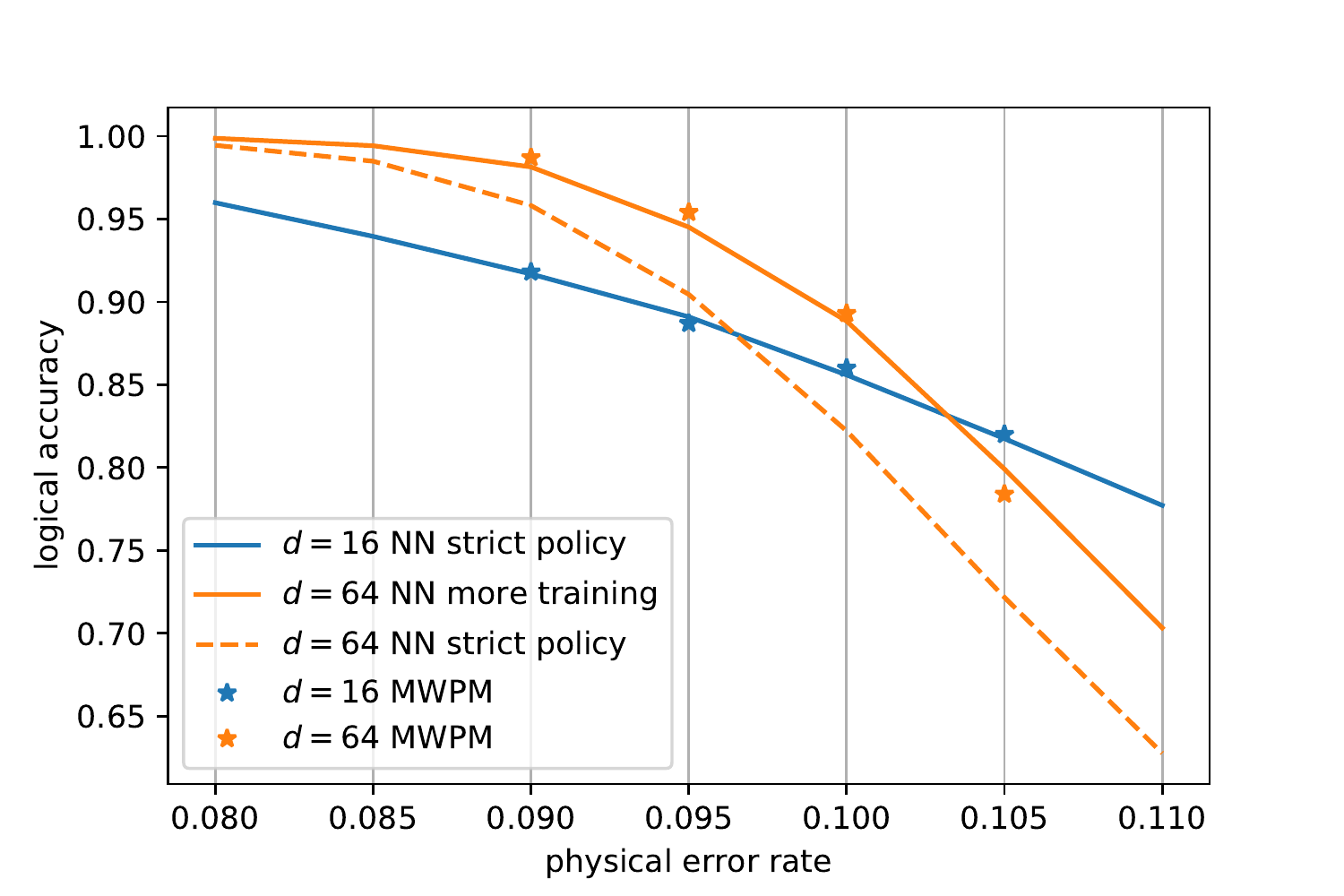}
	\caption{In this figure, we compare the performance of our neural decoders to the MWPM algorithm.
		The solid line for $d=16$ and the dashed line for $d=64$ decoder are using the strict training policy, while more training has been done on  the $d=64$ decoder corresponding to the solid line.
		The ``star'' points are the performance of the minimum-weight perfect matching algorithm.
		The colors of stars indicate the code distance they are evaluated on.
		The vertical grid indicates the physical error rates for which we evaluate the logical accuracy for the lines.
		We see the performance of neural decoders can be almost as good as MWPM for a decent range of physical error rates.
	}
	\label{fig:comparison_to_MWPM}
\end{figure}

For the strict training policy, we plot the logical accuracies versus the physical error rates in \autoref{fig:logical_vs_physical}.
Logical accuracy is simply $(1-\text{logical error rate})$ and is averaged over the two logical qubits.
For the solid lines, the decoders have been trained globally, i.e. have done both steps 1 and 2 in \autoref{subfig:complete_network}.
For the dashed lines, the decoders only did step 1, i.e. only the dense layers are trained.
The colors of the dashed lines indicate the code distance they are evaluated on.
The vertical grid indicates the physical error rates for which we evaluate the logical accuracy, where for each point we sample $10^4$ (syndrome, logical correction) pairs.
%The gray dotted line corresponding to $y=1-x$.
We can see that the solid lines cross around $p_{\text{physical}}=0.095$; therefore we might say our neural decoder has an effective threshold around $9.5\%$.
It can be seen that the global training is crucial for getting a decent performance because without it the effective threshold will be below $8\%$.

We can also spend more time to train the $d=64$ decoder, and then compare the performance of the neural decoders to the minimum-weight perfect matching algorithm (MWPM) in \autoref{fig:comparison_to_MWPM}.
The ``star'' points are the logical accuracies of MWPM, where each one is evaluated by $3000$ trials.
The $d=16$ decoder corresponding to the solid line and the $d=64$ decoder corresponding to the dashed line are from the strict training policy.
The $d=64$ decoder corresponding to the solid line is obtained by doing more training while having the same network architecture.
We see that without the strict training policy, the performance of the neural decoder is almost identical to MWPM for a decent range of physical error rates.
We can also compare to the renormalization group (RG) decoder in~\cite{duclos2010fast}, where the authors have shown a threshold of $8.2\%$ when using $2\times 1$ unit cell, and claim a threshold around $9.0\%$ if using $2\times 2$ unit cell.
With the strict training policy, our neural decoder is slightly better or at least comparable to the RG decoder, while without the policy our neural decoder is clearly better for $d\leq 64$.
\smallskip

\section{Discussion}
\label{sec:discussion}
One obvious question is whether we can get a good neural decoder for surface code or other topological codes on large lattices.
In the case of surface code, the major difference compared to the toric code is the existence of boundaries.
This means we have to inject some non-translational invariant components into the network.
For example, we can have a constant tensor $B$ with shape $(L,L,2)$ marks the boundary, e.g. $B(x, y, i)=1$ if $(x,y)$ is at the smooth boundary and $i=0$, or if $(x,y)$ is at the rough boundary and $i=1$; otherwise $B(x, y, i)=0$.
We then stack $B$ with the old input tensor before feed into the neural decoder.
More generally, if a renormalization group decoder exists for a topological code, we anticipate that a neural decoder can be trained to have similar or better performance.
For example, neural decoders for surface code with measurement errors, for topological codes with abelian anyons can be trained following the same procedure described in this paper.

Another question we want to discuss is the viability of our neural decoder at low physical error rates. On the one hand, we can train our neural decoders to approximate the RG decoder, and therefore they can have similar performance at low error rates. On the other hand, it will be much harder to improve neural decoders just by training on experimental data, because it will take a long time to encounter syndromes that are decoded incorrectly. Therefore, we should expect neural decoders to gradually lose the ability to adapt to experimental noise models as the physical error rates decrease.

We want to discuss a bit more about running neural networks on specialized chips.
It is straightforward to run our neural decoder on GPU or TPU~\cite{jouppi2017TPU} as they are supported by Tensorflow~\cite{tensorflow2015-whitepaper}, the neural network library used in this work.
There is software (e.g. OpenVINO) to compile common neural networks to run on commercially available field-programmable gate arrays (FPGAs), but we do not know how easy it is for our neural decoder\footnote{The only uncommon component of our neural decoder is element-wise parity function.}.
Apart from power efficiency, there is a study about operating FPGAs at 4K temperature~\cite{Lamb2016AnFPGAbased}.
Overall, there is a possibility to run neural decoders at low temperature.
Note that for running on FPGAs or benchmarking the speed, it is likely a good idea to first compress the neural networks, see~\cite{Cheng2018ModelCompression}.

\section{Acknowledgement}
\label{sec:acknowledgement}
The author wants to thank Ben Criger, Barbara Terhal, Thomas O'Brien for useful discussion.
The implementation of minimum-weight perfect matching algorithm, including the one used in \aref{appendix:varying_perror}, is provided by Christophe Vuillot, which uses the backend from either Blossom V~\cite{Kolmogorov2009BlossomV} or NetworkX~\cite{networkx}.
The author acknowledge support through the ERC Consolidator Grant No. 682726.

\bibliographystyle{unsrtnat}
\bibliography{main.bbl}

\appendix

\section{Implementation of Belief Propagation Algorithm}
\label{appendix:bp}
Belief propagation is a heuristic procedure for computing marginal probabilities of graphical models.
We choose to use a slightly different belief propagation implementation compared to~\cite{duclos2010fast}, as ours seems to be more natural for the bit-flip noise model.
We divide the lattice into $2\times 2$ unit cells.
Let $G$ be a bipartite graph, where one part corresponds to unit cells, and another part to coarse-grained edges.
Two vertices in $G$ is connected when the coarse-grained edge is adjacent to the unit cell.
This later decides how the messages flow in the graph.
However, we assign two variables $\{x(e_i),x(e_j)\}$ to each vertex corresponding a coarse-grained edge which $e_i$ and $e_j$ form.
In this section, the symbol $e$ or $e_i$ will denote original edges of the lattice (i.e. not coarse-grained).
We define $E$ to be the set of all edges $e$, $E_{cg} \subset E$ to be the set of $e$ which are components of coarse-grained edges (i.e. red edges in \autoref{fig:lattice}), and $\bar{E}_{cg} = E \setminus E_{cg}$.
Given a syndrome $S$, the unnormalized probability of an error configuration $\vec{x}\equiv \{x(e)\}_{e\in E}$ can be written as
\begin{equation}
p(\vec{x}) = g(S, \vec{x})\prod_{e\in E} p_e(x(e)) ,
\end{equation}
where $g(S, \vec{x}) = 1$ if $\vec{x}$ has syndrome $S$, and otherwise $g(S, \vec{x}) = 0$.
It is obvious $g(S, \vec{x})$ can be factorized to local terms.
Thus, the marginal distribution for $e\in E_{cg}$ can then be factorized according to $G$ as
\begin{equation}
\sum_{\{x(e), e\in \bar{E}_{cg}\}}p(\vec{x}) =  \prod_c f_c(\{x(e)\}_c) ,
\end{equation}
where the product is taken over all unit cells $c$, and $\{x(e)\}_c$ are the set of $x(e)$ such that $e\in E_{cg}$ is adjacent to $c$.
We can then apply the standard belief propagation to the graph $G$.
Without further explanation, we choose to use the following rule.
A unit cell $c_k$ sends to each of its adjacent cell $c_n$ a message containing 4 real numbers
\begin{equation}
\{m_{c_k, c_n}(x(e_i), x(e_j))\} \quad \text{for} \  x(e_i),x(e_j) = 0,1 ,
\end{equation}
where $e_i$ and $e_j$ form the coarse-grained edge between $c_k$ and $c_n$.
When we already fix an error configuration $\vec{x}$, we may use the simplified notations
\begin{equation}
p_e \equiv p_e(x(e)), \qquad m_{c_k, c_n} \equiv m_{c_k, c_n}(x(e_i), x(e_j)).
\end{equation}
To compute an out-going message from cell $c$, we take messages from the other three directions, and consider them to be the probability of error configuration on respective edges.
We then sum over all error configurations in the cell, which give the correct syndrome of the 4 plaquette stabilizer checks.
\begin{figure}
	\centering
	\includegraphics[width=0.5\linewidth]{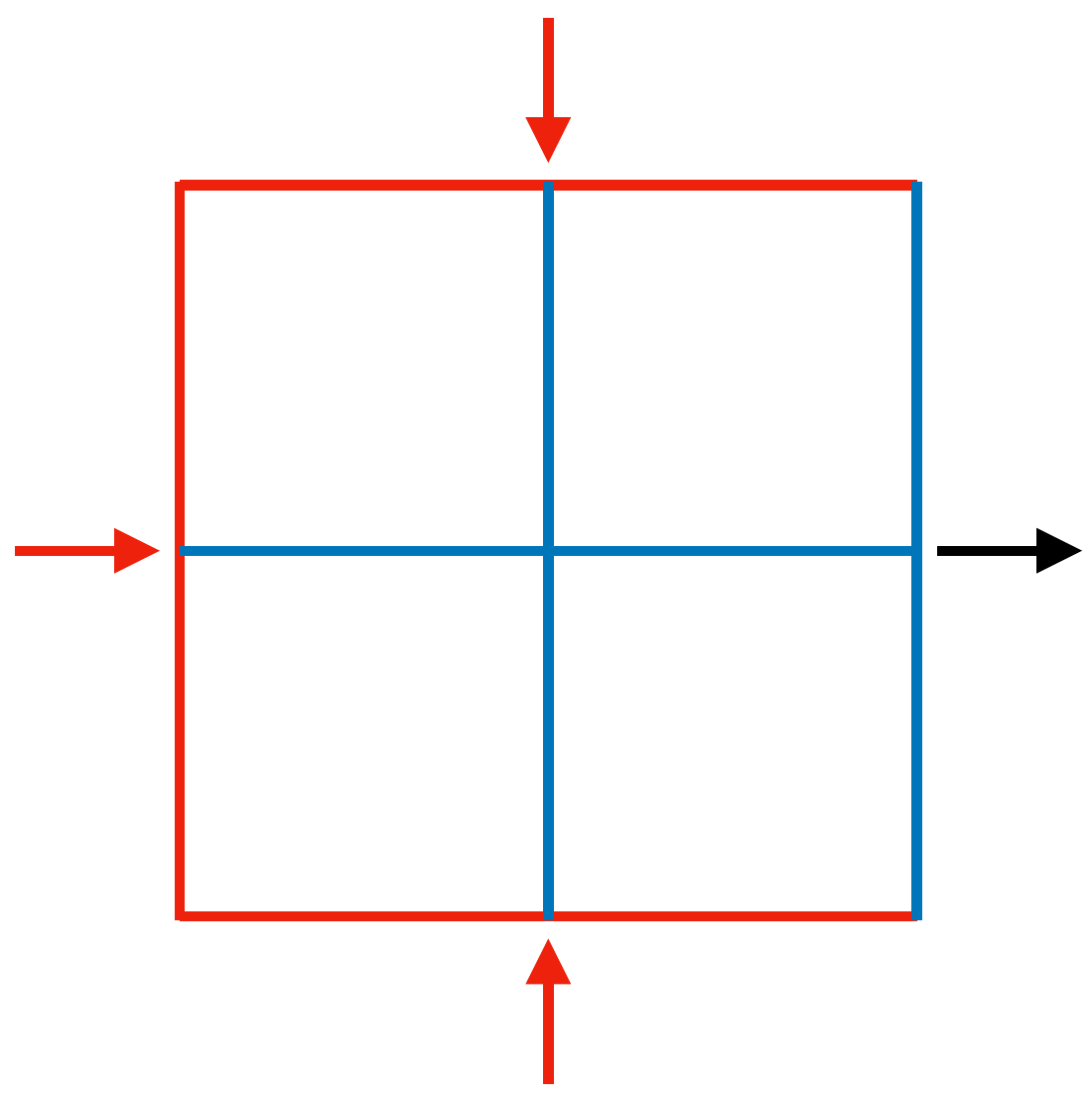}
	\caption{An illustration of message passing for a unit cell.}
	\label{fig:bp_cell}
\end{figure}
More concretely, we define $\vec{x}_c$ to be $\vec{x}$ restricted to edges in $c$ (i.e. all edges in \autoref{fig:bp_cell}), and $g'(S, \vec{x}_c)$ checks whether $\vec{x}_c$ is compatible with $S$ similar to $g$.
We assume we want to send the messages from $c$ to $c_n$, while the incoming messages are from cells $I = \{c_k\}$.
Then we have
\begin{equation}
m_{c, c_n} = \sum_{\vec{x}_c} g'(S, \vec{x}_c)\prod_{c_k\in I} m_{c_k, c} \prod p_{e_i}.
\end{equation}
For the last term $\prod p_{e_i}$, the product is taken over the blue edges in \autoref{fig:bp_cell}, assuming $c_n$ is on the right of $c$.

In the end of the message passing, we can compute the marginal probability by
\begin{equation}
P(x(e_i),x(e_j)) = m_{c_k, c_n}m_{c_n,c_k} / \left(p_{e_i}p_{e_j}\right) ,
\end{equation}
where $e_i$ and $e_j$ are the edges between $c_n$ and $c_k$.
From the joint distribution $P(x(e_i),x(e_j))$ we can compute the distribution $P\left(x(e_i)+x(e_j)\mod 2\right)$.
It is not hard to see that the above message passing rules will lead to the correct marginal probability when the underlying graph is a tree
(note this is not the case for the square lattices we are considering).
To generate training data for neural networks, we do 7 rounds of message passing defined above.

The key differences between our implementation and the one in~\cite{duclos2013fault} is
\begin{itemize}
	\item Ours utilizes all four stabilizer checks in each unit cell while in~\cite{duclos2013fault} only three are used.
	\item Each message contains 4 real numbers in our implementation while only 1 in~\cite{duclos2013fault}.
\end{itemize}

\section{Introduction to Neural Networks}
\label{appendix:intro_nn}
A neural network, at the highest abstraction, can just be viewed as a black-box function $f_{\text{nn}}(x,\vec{w})$ with many parameters $\vec{w}$ to be tuned.
We want $f$ to describe the input-output relation presented in a dataset $D=\{(x_i, y_i)\}$.
To do this\footnote{In this work, we choose to not study the effect of overfitting, as we have the ability to generate infinite data.}, we choose a (smooth) loss function $\mathcal{L}$, and then we do the minimization
\begin{equation}
\min_{\vec{w}}\sum_i \mathcal{L}(f(x_i, \vec{w}), y_i) .
\end{equation}
One important requirement is that $f$ is (almost-everywhere) differentiable with respect to $\vec{w}$.
This allows us to train the network with gradient descent, for which a good introduction can be found in~\cite{goh2017why}.
In general, we can expect gradient descent will take us to a local minimum or some region with very small gradients.
This is the advantage of ``end-to-end'' training compared to human-written heuristic algorithms, as the latter are unlikely to be a local optimum (assuming we can add real number parameters to those heuristic algorithms).
A common loss function for classification problems is cross-entropy.
Assume we have a dataset $D=\{(x_i, y_i)\}$ where $y_i\in \{0,1\}$, and the neural network output $y_i'$ which tries to approximate $\text{Prob}(y_i=1)$, the cross-entropy loss function is then calculated as following:
\begin{equation}
-\sum_i \left( y_i \log y_i' + \left(1-y_i\right) \log \left(1-y_i'\right)\right).
\end{equation}
Note that when we use the notation $\text{Prob}(y_i=1)$, we implicitly assume $D$ is obtained by sampling from an underlying probability distribution.

More concretely, most neural networks consist of many layers.
In this paper, the two relevant types of layers are the dense and convolution layer.
Dense layers (a.k.a fully-connected layers) have the form $g(A\vec{x}+\vec{b})$, where $g$ is some non-linear function applied entrywise, and the matrix $A$, vector $\vec{b}$ are the trainable parameters.
Assuming $A$ has a shape of $n\times m$, we will say the output dimension of the dense layer is $n$.
One convolution layer, as the name suggests, contains a collection of discrete convolutions.
For this paper, the input to the layer resides on a $2$-dimensional lattice of size $l^2$ with periodic boundary condition.
On each lattice site, there is a $d$-dimensional input vector $x_{\vec{u}} \in \real^d$, where the subscript $\vec{u} \in \mathbb{Z}_l^2$ (we use $\mathbb{Z}_l$ to denote integer in range $[0,l-1]$).
We define the kernel to be a tensor $K_{\vec{u},i}$, where $\vec{u} \in \mathbb{Z}_n^2$ and $i \in \mathbb{Z}_d$. With a slight abuse of notation, we will say such a kernel has size $n^2$. The convolution is then
\begin{equation}
\label{eq:convolution}
y_{\vec{v}} = \sum_{\vec{u} \in \mathbb{Z}_n^2} \sum_{i \in \mathbb{Z}_d} x_{\vec{v}-\vec{u}, i} K_{\vec{u}, i},
\end{equation}
where $x_{\vec{v}-\vec{u},i}$ is the $i$th element of $x_{\vec{v}-\vec{u}}$, and $\vec{v}-\vec{u}$ is calculated module $l$ because of the periodic boundary condition.
We will have a collection of kernels $\{K_{\vec{u}, i,j}\}_j$ for one convolution layer, which means we also have a collection of outputs $\{y_{\vec{v},j}\}_j$.
The cardinality of $\{K_{\vec{u}, i,j}\}_j$ is conventionally called the number of filters.
After \autoref{eq:convolution}, we can also apply a non-linear function $g$ entrywise if needed.

Before concluding this section, let us make one clarification.
In this paper, sometimes we only train part of the network, e.g. for blue circle 1 in \autoref{subfig:complete_network} we only train the dense layers.
Assuming $\vec{w}_1$ are the parameters in the part of the network we want to train and $\vec{w}_2$ are the rest, then we are doing the optimization
\begin{equation}
\min_{\vec{w}_1}\sum_i \mathcal{L}(f(x_i, \vec{w}_1, \vec{w}_2), y_i)
\end{equation}
by using some gradient descent optimizer.

\section{Comparison to Simpler Approachers}
\label{appendix:simpler_approach}
In this section, we will show the performance of the neural net decoders when trained with simpler approaches (more precisely, approaches with less human involvement and prior knowledge of toric code decoding), and provide some reasoning if possible.
The neural nets will be the same as the ones we used in the main text, except that they do not contain the ``removing complexity'' step in the post-processing.
We will see in general the performance gets much worse, especially when the lattice size grows large.
However, this does not mean these simpler approaches will always fail.
It just implies that a large amount of training time / human involvement is needed, which could make them impossible in practice.

The simplest approach is to train the whole network with input-output pairs (syndrome, logical correction).
During limited attempts, this approach does not produce decoders much better than random guess for large toric codes.
A hand-waving explanation is the following.
It is fair to assume a lot of parity functions need to be evaluated during the decoding process.
It is known that the parity is not an easy function for neural nets to compute~\cite{bengio2007scaling}, and one good way to approximate it is to increase the depth of the network.
So let us assume each renormalization stage needs 5 layers.
This means to decode $L=32$ toric code, the network will have 25 layers, which exceeds the range where neural nets can be reliably trained.

The problem of too many layers can be alleviated if we can pre-train the earlier layers of the network.
A similar strategy was used in training neural nets for computer vision problems~\cite{erhan2010does}.
For the bit-flip noise model, we can pre-train the earlier layers to mimic the renormalization group decoder.
Recall in \autoref{sec:toric_code_and_rn_decoder}, we mentioned that for the renormalization group decoder, the output corresponding to a coarse-grained edge $e$ is $p_e$.
Since we generate syndromes by first sampling $x(e)$ for all (not coarse-grained) edges $e$, we also have the ability to generate pairs (syndrome, $\{x(e)\}$ for coarse-grained edges $e$) for training.
Although $\{x(e)\}$ in the above pairs are binary numbers, with the cross-entropy as the cost function, in theory the output will converge to $p_e(1)$.
The pre-training is done one renormalization block at a time.
More concretely, with the network we are using in the main text, we will train the output of the 12th layer with the training target of first renormalization block, and the output of the 24th layer with the second block, etc.
We can try this method on $L=32$ toric code and bit-flip error rate $0.08$.
It does not work well, as we can see in \autoref{tab:cross_entropy_after_blocks}.
\begin{table}
	\begin{tabular}{c c c c c c}
		Block & RN1 & RN2 & RN3 & RN4 & Dense\\
		Cross-entropy  & 0.16 & 0.22 & 0.28 & 0.4 & 0.5 \\
		Accuracy & 0.93 & 0.90 & 0.86 & 0.79 & 0.58
	\end{tabular}
	\caption{Cross-entropy and accuracy after each renormalization block when we do the training block by block.
		It is done with a bit-flip error rate $0.08$.
		Although the accuracy after the first RN block is comparable to the input error rate, it gradually decreases until barely above $0.5$}
	\label{tab:cross_entropy_after_blocks}
\end{table}
These numbers are not very accurate and coming from a single training instance.
However, the author has observed the same trend many times that the loss and accuracy slowly degrade in the process of renormalization, even though the error rate is way below the theoretical threshold.
To diagnose the reason, we first notice that the first RN block actually performs reasonably well.
This suggests that the optimizer is capable of training each RN block alone.
Assume this is indeed the case, the degrading of performance is likely caused by the following two reasons.

First, later in the renormalization process, if we look at the coarse-grained syndrome or $p_e$ alone, they behave more and more like white noise.
While it is possible for the neural nets to do the same post-processing described in \autoref{subsec:design_network}, a few layers of the network will be occupied by this.
Therefore, the natural solution is to implement the post-processing ourselves.
By doing this, we suspect it is possible to reach a threshold of $8\%$, but apparently $8\%$ is still not good enough.

The second reason is related to the convergence of $p_e$.
When below the threshold, we will encounter very often that $p_e$ is very close to $0$ or $1$.
For example, if $x(e)=1$ vs $x(e)=0$ corresponding to a weight $4$ vs weight $1$ local configuration, then we will have $p_e \approx p_0^3$, where $p_0$ is the initial error rate.
This will become more prominent later in the renormalization process, as $p_e$ from the previous renormalization stage become the error rate in the next stage.
It is important to know how close to $0$ (or $1$) $p_e$ is on the logarithmic scale.
Otherwise, in the later stage, the information will not be accurate enough to deduce configuration close to the minimum weight of errors.
This poses the following requirements:
\begin{itemize}
	\item When we pass $p_e$ to some intermediate layer of a neural net, it should be able to distinguish between small $p_e$.
	However, recall that each layer does the computation $f(Ax + b)$, where $f$ has a bounded derivative.
	Thus, to distinguish a set of small $\{p_e\}$, we need $\| A \| \sim 1/\min \{p_e\}$.
	This will either not be achieved by training or cause instability of the network.
	Another issue related to the minuscule nature of $p_e$ is the cross-entropy loss function does not provide enough motivation for $p_e$ to converge to the target value $q$ in log-scale.
	More accurately, the derivative of the cross-entropy scales like $O(|p_e - q|)$ when $p_e \approx q$, which will be too small before the convergence in log-scale.
	A natural solution is we replace the appearance of $p_e$ with $\log p_e -\log (1-p_e)$.
	\item Even with a good representation of $p_e$, we shall still be very cautious about the training, as we are trying to estimate very small $p_e$ from sampling.
	In the end, we decide to implement a belief propagation routine and use the input-output pair from the routine to train the network.
	The advantage is that belief propagation directly outputs the probability, which should be reasonably accurate in the logarithmic scale.
	Therefore, we can get a much stable training process.
\end{itemize}

\section{Technical Details}
\label{appendix:details}
The objective of this section is to describe some technical details for people who do not plan to read the source code.

For the majority of network, we use leaky ReLUs~\cite{Maas2013Rectifier} as the activation function, which has the form
\begin{equation}
y=x \quad \text{if} \  x>0; \quad y=0.2x \quad \text{if} \  x\leq0
\end{equation}
Apart from the last layer of each renormalization block and the last dense layer, the number of filters in each convolution layer is 200, and the output dimension of each dense layer is 50.

For training a belief propagation network, we generate a dataset of size $80000$.
The dataset consists of the input and output of the belief propagation algorithm described in \aref{appendix:bp} when applied to $d=16$ toric code.
The optimizer we use is the ADAM~\cite{kingma2014adam}.
The learning rate parameter of the optimizer is set to $7\times 10^{-4}$ for training belief propagation network, $10^{-3}$ for training dense layers, $7\times 10^{-5}$ for global training of $L=16,32$ lattice, and $7\times 10^{-6}$ for $L=64$.
The batch size for training is 50.
The training of the dense layers uses around 1000 batches.
For the strong policy, the global training uses 3000 batches regardless of the code distance.
To see the potential of our neural decoder at $d=64$, we also did a longer training and compared it to MWPM in \autoref{fig:comparison_to_MWPM}.
In total, it is trained using 18000 batches.
The first 12000 batches are trained on physical error rate $p=0.09$, and the last 6000 batches are on $p=0.095$.
The reason of switching to a higher error rate for late-stage training is that the accuracy at $p=0.09$ is too close to 1 for effective training.

%ADAM keeps track of t by saving $beta^t$.
%Batch normalization moving mean is trainable in Tensorflow.

\section{Spatially Varying Error Rates}
\label{appendix:varying_perror}
A natural use of the neural decoder is to train it with experimental data.
Naturally, the noise models in the experiments will not be translational invariant.
There are two simple ways to reconcile this fact with the translational-invariance of convolutional neural nets:
\begin{enumerate}
	\item Allow the first few layers of the network to be non-translational invariant.
	\item Introduce site-dependent trainable variables to the networks.
\end{enumerate}
In this section, we will consider the error model that has varying bit-flip error rates across the lattice.
For this, we can use the second approach, where the site-dependent variables can in principle represents the varying error rates.
In fact, recall that the neural decoder has an input shape $(L,L,3)$, which contains $2L^2$ numbers that are originally error rates feeding into belief propagation.
Thus we can simply feed the site-dependent variables into the neural decoder and then train them.

However, there is still a complication regarding the starting point of this training.
A natural choice is to start with the trained neural decoder for uniform error rate and only train the site-dependent variables.
With this route, there is a risk that if previously we trained the neural decoder under uniform error rate for long enough, the neural decoder could learn to ignore the error rate inputs as they are constant.
In this case, only training the site-dependent variables can fail.

Another route we can take is that we also reinitialize the first renormalization block.
More accurately, we do the following:
\begin{enumerate}
	\item We start with the trained neural decoder for uniform error rate.
	\item We reverse the belief propagation network in the first renormalization block to the pre-trained weights. In other words, it now again approximates belief propagation.
	\item We then train the site-dependent variables and the first renormalization block together.
\end{enumerate}
We test this scheme with a distance 16 toric code.
For each qubit, there is a $50\%$ chance it has bit-flip rate $0.16$, and a $50\%$ chance of rate $0$.
In other words, only around half of the qubits are noisy.
With this new error model, we train the site-dependent variables and the first renormalization block for 4500 batches, where each batch contains 50 training data.
The learning rate of the ADAM optimizer is set to $2\times 10^{-4}$.
By performing this training, we increase the logical accuracy from $0.967$ to $0.993$, where $0.967$ corresponds to the decoder trained on uniform bit-flip error model.
The accuracy is each evaluated by using $10^5$ (syndrome, logical correction) pair.
To provide some comparison, we run the minimum-weight perfect matching (MWPM) algorithm for the same error model.
Without providing error rate information, the MWPM algorithm assigns an equal weight to each qubit, and have a logical accuracy of $0.975$.
If we provide the perfect information about the error rates, we can assign weight 1 to noisy qubits, and weight 100 to the noiseless ones.
For each pair of violated parity checks, we can select the path with the minimum total weight between them, and use this weight for the MWPM algorithm.
With this choice, the logical accuracy rises to $1$, e.g. $100\%$ success.
These accuracies are each evaluated by $10^4$ (syndrome, logical correction) pair.

Based on the thoughts above, it is likely better to not start with the neural decoder trained on uniform error rate.
Instead, we can train a neural decoder with training data that has varying error rates, but otherwise the same procedure as depicted in \autoref{subfig:complete_network}.
This way, the first renormalization block will not learn to ignore the error rate inputs.

\end{document}